\documentclass[showpacs,aps,pra,twocolumn,10pt,superscriptaddress]{revtex4-2}
\usepackage{amsmath,graphicx,color}

\begin{document}

\title{Nonlinear dynamics in the balanced two-photon Dicke model with qubit dissipation}

\author{Jiahui Li}
\affiliation{Beijing Computational Science Research Center, Beijing 100193, China}

\author{Stefano Chesi}
\email{stefano.chesi@csrc.ac.cn}
\affiliation{Beijing Computational Science Research Center, Beijing 100193, China}
\affiliation{Department of Physics, Beijing Normal University, Beijing 100875, People’s Republic of China}

\begin{abstract}

We study the complex nonlinear dynamics of the two-photon Dicke model in the semiclassical limit by considering cavity and qubit dissipation. In addition to the normal and superradiant phases, another phase that contains abundant chaos-related phenomena is found under balanced rotating and counter-rotating couplings. In particular, chaos may manifest itself through period-doubling bifurcation, intermittent chaos, or quasi-periodic oscillation, depending on the value of qubit frequency. Transition mechanisms that exist in these three distinct routes are investigated through the system's long-time evolution and bifurcation diagram. Additionally, we provide a comprehensive phase diagram detailing both the existence of stable fixed points and the aforementioned chaos-related dynamics.

\end{abstract}

\maketitle

\emph{\textbf{Keywords:}}
{two-photon; qubit dissipation; nonlinear dynamics; chaos}

\section{introduction}
\label{introduction}

The quantum Dicke model, describing a collective interaction between $N$ qubits and a single bosonic mode, has sparked profound research in the past decades \cite{1954Coherence,kirton2019introduction}. In the strong coupling regime, the collective coupling can induce many interesting non-equilibrium phenomena, such as a superradiant phase (SP) transition \cite{Klaus1973Equilibrium,dimer2007proposed,Hwang2015Universal,dalla2016dicke,larson2017some,fitzpatrick2017observation,kessler2012dissipative,liu2017universal,2017Suppressing,hwang2018dissipative}, multi-stability \cite{gelhausen2017many,gelhausen2018dissipative}, classical and quantum chaos \cite{emary2003chaos,emary2003quantum,hou2004decoherence,lambert2009quantum,perez2011excited}. The implementation of this model with analog quantum simulators makes it possible to examine these phenomena in several experimental platforms \cite{bishop1996application,ashhab2010qubit}, such as trapped ions \cite{puebla2016robust,lv2018quantum}, arrays of NV centers \cite{zou2014implementation}, and superconducting circuits \cite{beaudoin2011dissipation,lamata2017digital,zhang2018generalized}. In these driven-dissipative systems, multi-photon processes can be selected from linear interactions by properly adjusting the driving lasers' parameters. An interesting variation of the Dicke model that considers a two-photon interaction has received much interest recently, as it presents a variety of intriguing dynamics behaviors \cite{travvenec2012solvability,puebla2017protected,cheng2018nonlinear,felicetti2018two,maldonado2019squeezed}. For example, there exists a critical coupling at which a second-order transition to a superradiant-like phase may occur \cite{garbe2017superradiant,chen2018finite,xie2019generalized,cui2019two}. At variance with the one-photon model, this superradiant phase transition is associated with a squeezed cavity field, thus has relevance for research on enhanced photon squeezing \cite{banerjee2022enhanced}. In the ultra-strong coupling limit, the two-photon Dicke model undergoes a spectral instability, when discrete energy levels collapse into a continuous band \cite{felicetti2015spectral,duan2016two,cong2019polaron}. This dynamical property is preserved in the presence of bosonic field dissipation \cite{li2022nonlinear}. 

Furthermore, the occurrence of complex nonlinear dynamics and chaos, in particular, is important for several areas of classical and quantum physics \cite{chavez2016classical,zhu2022cavity}, as well as for Dicke-type models which allow for semiclassical approximation in the large $N$ limit \cite{patra2019driven,patra2019chaotic,lerma2019dynamical}. When neglecting the influence of the environment, a transition from quasi-integrability to quantum chaos is predicted to occur after the superradiant phase transition has taken place  \cite{buijsman2017nonergodicity,chavez2019quantum,wang2019effect,wang2020statistical}. Going beyond closed-system evolution, the observation of classical chaos, has been reported in models with an unbalanced interaction, i.e., where the coupling strength of the counter-rotating terms is larger than the co-rotating terms \cite{kirton2018superradiant,stitely2020nonlinear,li2022nonlinear}. Taking cavity loss into consideration, \cite{li2022nonlinear} and   \cite{stitely2020nonlinear} discussed nonlinear dynamics of the one- and two-photon Dicke models respectively, finding a rich dynamical behavior including Hopf bifurcations, period-doubling, and strange attractors. While these studies assumed that the dominant decoherence mechanism is from the bosonic field, the influence of qubit relaxation and dephasing on the system's long-time evolution has not been characterized in detail.

In light of these considerations, we present in this paper a study of nonlinear dynamics in the two-photon Dicke model, including the influence of bosonic and atomic dissipation. We find that qubit decoherence has a profound influence on the systems properties, leading to qualitative changes in the nonlinar behavior: Not only it benefits system stability but also leads to a richer scenario for the occurrence of chaotic dynamics. Differently than models without qubit decoherence \cite{li2022nonlinear,stitely2020nonlinear}, additional routes to chaos are available (besides period-doubling bifurcations) and the chaotic dynamics does not rely on strong counter-rotating terms. Therefore, in the following we will only consider the isotropic two-photon Dicke model, with balanced interactions. By restricting ourselves to the isotropic limit, our study can be also understood as a detailed analysis of the 'instable phase' (I), recently revealed by a study of stable fixed points~\cite{garbe2020dissipation}.

The paper is organized as follows. In Sec.~\ref{model}, we describe the open two-photon Dicke model and explore its phase diagram in the mean-filed approximation, only considering stable fixed points. In Sec.~\ref{routes}, we investigate the long-time nonlinear dynamics and find three distinct routes to chaos, depending on the qubit frequency: (i) period-doubling; (ii) quasi-periodicity; (iii) intermittency; In Sec.~\ref{phase}, we present the overall phase diagram, involving stable fixed points as well as chaos-related nonlinear dynamics. A comparison of our results with previous relevant research can also be found. Finally, we provide a summary of our findings in Sec.~\ref{conclusion}.

%%%%%%%%%%%%%%%%%%%%%=================The model================%%%%%%%%%%%%%%%%%%%%%
\section{The model}
\label{model}
%%%%%%%%%%%%%%%%%%%%%%%%%%%%%%%%%%%%%%%%%%%%%%%%%%%%%%%%%%%%%%%%%%%%%%%%%%%

We consider a two-photon Dicke model described by the Hamiltonian ~\cite{garbe2020dissipation}
\begin{align}
\label{eqH}
\hat{H}=\omega_0\hat{a}^{\dagger}\hat{a}+\frac{\omega_q}{2}\sum_{j=1}^{N}\hat{\sigma}_z^{(j)}+\frac{g}{\sqrt{N}}\sum_{j=1}^{N}\hat{\sigma}_x^{(j)}(\hat{a}^2+\hat{a}^{\dagger 2}).
\end{align}
where $\hat{a}~(\hat{a}^{\dagger})$ is the annihilation (creation) operator of a bosonic field of frequency $\omega_0$ and $\vec{\hat{\sigma}}^{(j)}$ are the Pauli operators of $N$ qubits. The qubits have a uniform transition frequency $\omega_q$  and coupling strength $g$ with the bosonic mode. As a consequence, Eq.~(\ref{eqH}) can be rewritten in terms of collective angular momentum operators $\vec{\hat{J}}=\frac{1}{2}\sum_{j=1}^{N}\vec{\hat{\sigma}}^{(j)}$ as follows: 
\begin{align}
\label{eqH2}
\hat{H}=\omega_0\hat{a}^{\dagger}\hat{a}+\omega_q\hat{J}_z
+\frac{2g}{\sqrt{N}}\hat{X}\hat{J}_x.
\end{align}
Here we have also defined $\hat{X}=\hat{a}^2+\hat{a}^{\dagger2},\hat{Y}=i(\hat{a}^2-\hat{a}^{\dagger2})$. The two-photon Dicke Hamiltonian is invariant under the generalized parity operator $\hat\Pi=(-1)^N\bigotimes_{j=1}^{N}\hat{\sigma}_z^{(j)}e^{i\pi\hat{a}^{\dagger}\hat{a}/2}$~\cite{felicetti2015spectral}, leading to
\begin{align}
\label{exchange}
\hat{a}\rightarrow i\hat{a},\,\,\, \hat{\sigma}_{x,y}\rightarrow -\hat{\sigma}_{x,y}.
\end{align}
As a result, the system features a four-fold symmetry. 

The two-photon Dicke model is well-known for its spectral properties, with the discrete eigenenergy levels collapsing into a continuous band in the ultra-strong coupling regime \cite{felicetti2015spectral,duan2016two,cong2019polaron}. Cavity field dissipation alone cannot eliminate this instability, as a diverging photon number is obtained when approaching the corresponding `localized' phase \cite{li2022nonlinear}. Before this instability a superradiant-like phase transition takes place, in which the collective pseudospin attains a macroscopic mean value and the bosonic field is driven to a squeezed state \cite{garbe2017superradiant,chen2018finite,banerjee2022enhanced}. As we will see, however, the properties of the model depend sensitively on the dissipative environment. Here, as in Ref.~\onlinecite{garbe2020dissipation}, we include all three dissipation channels which naturally appear in the system: photon loss, individual qubit decay, and qubit dephasing. We describe the system evolution with the standard master equation:
\begin{align}
\label{eq-ME} 
\dot{\hat{\rho}}=-i[\hat{H},\hat{\rho}]+\kappa\mathcal{D}[\hat{a}]\hat{\rho}+\sum_{j=1}^{N}\left(\Gamma_{\downarrow}\mathcal{D}[\hat{\sigma}_-^{(j)}]\hat{\rho}+\Gamma_{\phi}\mathcal{D}[\hat{\sigma}_z^{(j)}]\hat{\rho}\right),
\end{align}
where the Lindblad superoperators are defined as $\mathcal{D}[\hat{A}]\hat{\rho}=2\hat{A}\rho \hat{A}^{\dagger}-\hat{A}^{\dagger}\hat{A}\rho-\rho \hat{A}^{\dagger}\hat{A}$. The Lindblad master equation is left unchanged by the transformation in Eq.~(\ref{exchange}) [i.e., $\hat\Pi^\dag \hat{\rho} \hat\Pi$ is also a solution of Eq.~(\ref{eq-ME})], thus retains the four-fold symmetry of the model. 

In this work, we mainly focus on the limit of large $N$ and consider coupling strengths comparable to the bosonic-field frequency, when the spin degrees of freedom acquire macroscopic populations. This justifies applying the mean-field approximation to the system's dynamics. Decoupling cavity-qubit correlations as $\langle \hat{C}\hat{Q} \rangle = \langle \hat{C}\rangle \langle\hat{Q} \rangle$,  the following equations of motion are obtained:
\begin{align}
\label{eq-MFfirst}
\frac{d \langle \hat{X} \rangle }{dt}&=-2\kappa\langle \hat{X} \rangle  -2\omega_0 \langle \hat{Y} \rangle ,\\
\frac{d  \langle \hat{Y} \rangle }{dt}&=-2\kappa\langle \hat{Y} \rangle +2\omega_0 \langle \hat{X} \rangle  +4g\sqrt{N}s_x\left(2 \langle \hat{a}^{\dagger}\hat{a}\rangle+1\right),\\
\frac{d  \langle \hat{a}^{\dagger}\hat{a}\rangle  }{dt}&=-2\kappa  \langle \hat{a}^{\dagger}\hat{a}\rangle+2g\sqrt{N}s_x \langle\hat{Y} \rangle ,\\
\frac{d s_x }{dt}&=-\omega_q s_y-2\Gamma's_x, \\
\frac{d s_y }{dt}&=\omega_q s_x -\frac{2g}{\sqrt{N}}s_z \langle \hat{X} \rangle  -2\Gamma's_y, \\
\label{eq-MFlast}
\frac{d s_z }{dt}&=\frac{2g}{\sqrt{N}}\langle \hat{X} \rangle s_y -2\Gamma_{\downarrow}(s_z +1),
\end{align}
where we rescaled the expectation values of the spin operators as $\vec{s}=\langle 2\vec{\hat{J}}\rangle/N$ and defined $\Gamma'=2\Gamma_{\phi}+\Gamma_{\downarrow}/2$. Due to the presence of qubit dissipation, the collective spin evolution will not be restricted to the unit sphere, giving a six-dimensional phase space (instead of the five-dimensional phase space of Ref.~\cite{li2022nonlinear}). 
 
As a first step of our investigation we focus on the stationary states, whose stability can be established from the Jacobian matrix and the Routh-Hurwitz criteria \cite{strogatz2018nonlinear,dejesus1987routh}. As expected, the system presents a phase transition from the normal phase ($\rm{NP}$) to the superradiant phase ($\rm{SP}$). The normal phase fixed point describes $s_z=-1$ spins accompanied by a zero photon number~\cite{garbe2020dissipation}:
\begin{align}
\label{eqNP}
{\rm{NP}}:\,\,\, s_z=-1, \langle \hat{a}^{\dagger}\hat{a}\rangle=0,
\end{align}
and is stable for ~\cite{garbe2020dissipation}:
\begin{align}
\label{eq-gt1}
g<g_{t1}=\sqrt{\frac{(\kappa^2+\omega_0^2)(\omega_q^2+4\Gamma^{'2})}{4\omega_0\omega_q}}.
\end{align}

\begin{figure}
\centering
\includegraphics[width=0.5\textwidth]{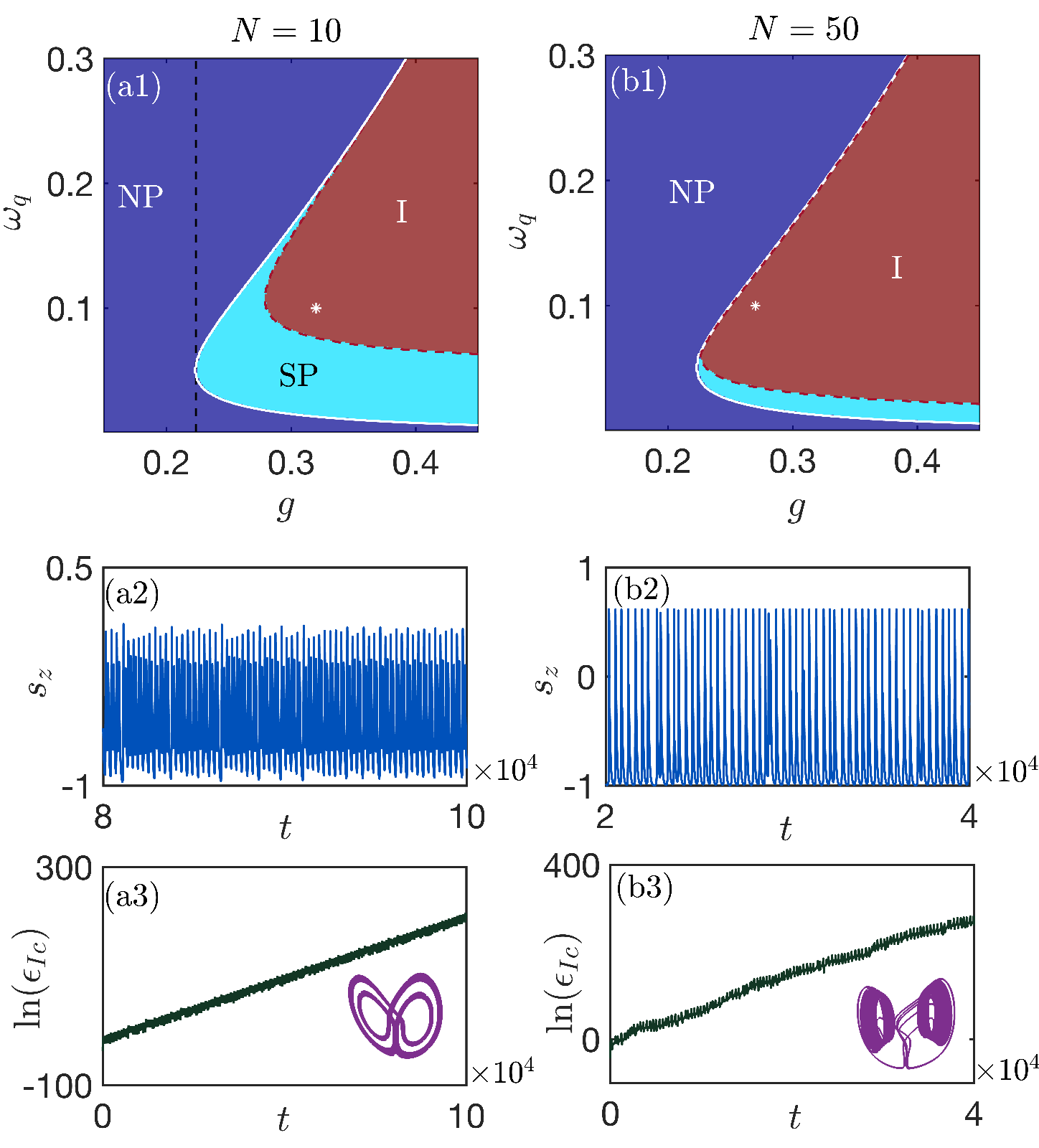}
\caption{\label{fig1} Panels (a1) and (a2): phase diagrams of stable fixed points for different $N$. The dark blue area represents the normal phase ($\rm{NP}$); the light blue area represents the superradiant phase ($\rm{SP}$), and the red area represents the phase without any stable fixed point ($\rm{I}$). Critical point $g_{t1}$ (write line) distinguish $\rm{NP}$ and other phases. $g_{t2}$ (red dashed line) is the boundary of $\rm{SP}$ and phase $\rm{I}$, which is obtained numerically. $g_u$ (black dashed line) is one of the critical points in the two-photon Dicke model without qubit decay \cite{li2022nonlinear}. It intersects with $g_{t1}$ at the knee occasionally in panel (a1) and out of the illustrated coordinate range in panel (b1). Panels (a2) and (a3): chaotic motion in the phase $\rm{I}$ for $ N=10, \omega_q=0.1, g=0.32$. Panels (b2) and (b3): chaotic motion in the phase $\rm{I}$ for $N=50, \omega_q=0.1, g=0.27$. Insets: corresponding qubit trajectories in phase space. The system parameters are in units of the cavity field frequency $\omega_0$: $\kappa=1$, $\Gamma_{\downarrow}=\Gamma_{\phi}=0.01$.
}
\end{figure}

In most literature, $1/N$ scaling in the interaction is adopted in the two-photon Dicke Hamiltonian, i.e., the coupling strength is written as $g/N$ \cite{felicetti2015spectral,garbe2017superradiant,chen2018finite,cui2019two,banerjee2022enhanced}. However, from the point of view of the normal state instability, it is quite interesting to consider the alternative scaling $1/\sqrt{N}$, which we have used in Eq.(\ref{eqH}). As seen in the phase diagrams of Fig.~\ref{fig1} (white solid curves), there is a well-defined phase boundary independent of $N$~\cite{garbe2020dissipation}. If $1/N$ scaling were adopted in Eq.~(\ref{eqH}), the stability boundary would differ by a factor $\sqrt{N}$ in $g_{t1}$, giving a critical line which depends on qubit number. Considering the thermodynamic limit $N\rightarrow\infty$ at fixed $g$, the normal state would always stable as the critical line will shift right with increasing $N$.  Therefore, in the following, we will restrict ourselves to the type of thermodynamic limit implied by Eq.~(\ref{eqH}), when the nontrivial instability of the normal phase remains independent of $N$.

%%%%%%%%%%%%%%%%%%%%============fig1===========================%%%%%%%%%%%%%%%%%%%
%%%%%%%%%%%%%%%%%%%%%%%%%%%%%%%%%%%%%%%%%%%%%%%%%%%%%%%%%%%%%%%%%%%%%%%%%%%
 
The second type of fixed point is a superradiant state with excited spins ($s_z>-1$) and a non-zero photon number:
\begin{align}
\label{eqSP}
{\rm{SP}}:\,\,\,&s_z  = -\frac{1+\beta}{2}+\sqrt{\left(\frac{1+\beta}{2}\right)^2-\beta\frac{g_{t1}^2}{g^2}}, \notag \\
&\langle \hat{a}^{\dagger}\hat{a}\rangle=\frac{1}{2\beta}(s_z+1),
\end{align}
where $\beta=\omega_0\Gamma'/(N\omega_q\Gamma_{\downarrow})$. As the system has a four-fold symmetry, the $\rm{SP}$ fixed points of Eq.~(\ref{eqSP}) appear in pairs, distinguished by the values of $\pm\langle \hat{X}\rangle, \pm\langle \hat{Y}\rangle, \pm s_{x,y}$.  They exist for $\Gamma_{\downarrow}>0$, otherwise $\rm{NP}$ is the only possible state of the system. In the parameter scope indicated in Fig.~\ref{fig1}, the SP states are stable when $g_{t1}<g<g_{t2}$. $g_{t2}$ is obtained numerically and plotted by the red dashed line in the figure. SP states also appear at large coupling strength which is beyond the parameter range considered in this work \cite{garbe2020dissipation}. Obviously, the SP phase will shrink with increasing qubit number.

We recall the phase diagram for stable fixed points in Fig.~\ref{fig1}(a) and (b), which is plotted under small qubit dissipation and isotropic couplings. Phase $\rm{NP}$ ($\rm{SP}$) indicates where the fixed point $\rm{NP}$ ($\rm{SP}$) is stable. The phase $\rm{I}$ denotes the absence of a stable fixed point. More details on phase diagram change with different system parameters can be seen in Ref.~\cite{garbe2020dissipation}, which investigates stationary dynamics of the two-photon Dicke model by the mean-field decoupling approximation. Considering the effects of both qubit and cavity dissipation, research indicates that, at isotropic coupling, there is a threshold of qubit dissipation ($\Gamma/\omega_0\approx0.38$) beyond which SP and bistable phases exist; otherwise, unstable phase I occupy the bulk of the phase diagram. While the study is limited to stationary states, which is unexpected that the unstable region conceals abundant nonlinear dynamics. As shown in Fig.~\ref{fig1}(a2)-(b3), both in small ($N=10$) and large ($N=50$) systems, we found chaotic motions in the unstable phase, as well as other chaos-related nonlinear behaviors. 

Fig.~\ref{fig1} (a) and (b) present a simple structure when compared to the phase diagram of the anisotropic two-photon Dicke model \cite{li2022nonlinear}, which considers imbalanced rotating and counter-rotating coupling. Taking only cavity decay into account, the system displays a localized phase $\rm{U_0}$ reflecting the spectral collapse of the closed-system Hamiltonian. A constant critical point $g_{u}=\sqrt{\kappa^2+\omega_0^2}/2$ distinguishes $\rm{U_0}$ from other phases, beyond that system tends to evolve to poles of the Bloch sphere with diverging photon numbers. At the same time, various coexistence phases lead to complicated and volatile phase diagrams, which depend on initial conditions. The localized phase and various coexistence phases vanish in Fig.~\ref{fig1}, which acknowledges the role of qubit dissipation in the stabilization of the two-photon Dicke model. Further, in Ref.~\cite{li2022nonlinear}, the anisotropic parameter $\lambda$ plays an important role in the nonlinear behaviors that chaos is only found after the pole-flip transition $\lambda>\lambda_t>1$. 

As an extension of the above studies, the current work is implemented in the isotropic condition $\lambda=1$ and will focus on the nonlinear dynamics in the unstable phase. Small qubit decay is clearly more advantageous for our simulation, which will provide a large area of unstable phase. $\Gamma_{\downarrow}=\Gamma_{\phi}=0.01$ are adopted in Fig.~\ref{fig1}(a) and (b), and will also be used throughout this work. Before presenting the main results, we noticed that system dynamics are inevitably influenced by qubit number, just the same as the above-mentioned studies. Though the cavity field does not exhibit a well-defined classical occupation in the $\rm{SP}$ phase, mean-field treatment is nevertheless applicable to the given system since the atomic field has a macroscopic population \cite{li2022nonlinear}. And it has little effect on $\rm{I}$ phase, which exhibits a macroscopic oscillation in the evolution of $\bar{n}_c(t)=\langle \hat{a}^{\dagger}\hat{a}\rangle(t)/N$. As shown in the insets of Fig.~\ref{fig4} (a3) and (b3), the photon field oscillation is about $0<\bar{n}_c(t)<8$ with samll qubit number $N=10$. Large fluctuations of photon field also pose higher requirements for our computing conditions, especially in larger systems. Since chaos can be found in both small and large systems, (actually easier to discovered in large system), we adopt $N = 10$ throughout our calculation to circumvent numerical difficulties caused by large qubit numbers.

%%%%%%%%%%%%%%%%%%%%%%%%%%%%%%%%%%%%%%%%%%%%%%%%%%%%%%%%%%%%%%%%%%%%%%%%%%%
\section{routes to chaos}
\label{routes}
%%%%%%%%%%%%%%%%%%%%%%%%%%%%%%%%%%%%%%%%%%%%%%%%%%%%%%%%%%%%%%%%%%%%%%%%%%%

Even though there is no stable fixed point in phase $\rm{I}$, it is  more complicated than it looks: We found various types of nonlinear dynamics in this phase, such as periodic, quasi-periodic, and chaotic motion. Generally, chaos can appear in a non-linear three (or higher) dimensional deterministic system via local bifurcation. In the discussed two-photon Dicke model, altering qubit frequency $\omega_q$ allows chaos to manifest in three distinct ways. In the following, we present a detailed analysis of the bifurcation processes of how the system enters into the chaotic regime by (i) period-doubling bifurcation, (ii) intermittency, and (iii) quasi-periodicity.

%%%%%%%%%%%%%%%%%%%%%%%%%%%%%%%%%%%%%%%%%%%%%%%%%%%%%%%%%%%%%%%%%%%%%%%%%%%
\subsection{ Period-Doubling bifurcation }
%%%%%%%%%%%%%%%%%%%%%%%%%%%%%%%%%%%%%%%%%%%%%%%%%%%%%%%%%%%%%%%%%%%%%%%%%%%
\label{period}

%%%%%%%%%%%%%%%%%%%%============fig2===========================%%%%%%%%%%%%%%%%%%%
\begin{figure}
\centering
\includegraphics[width=0.5\textwidth]{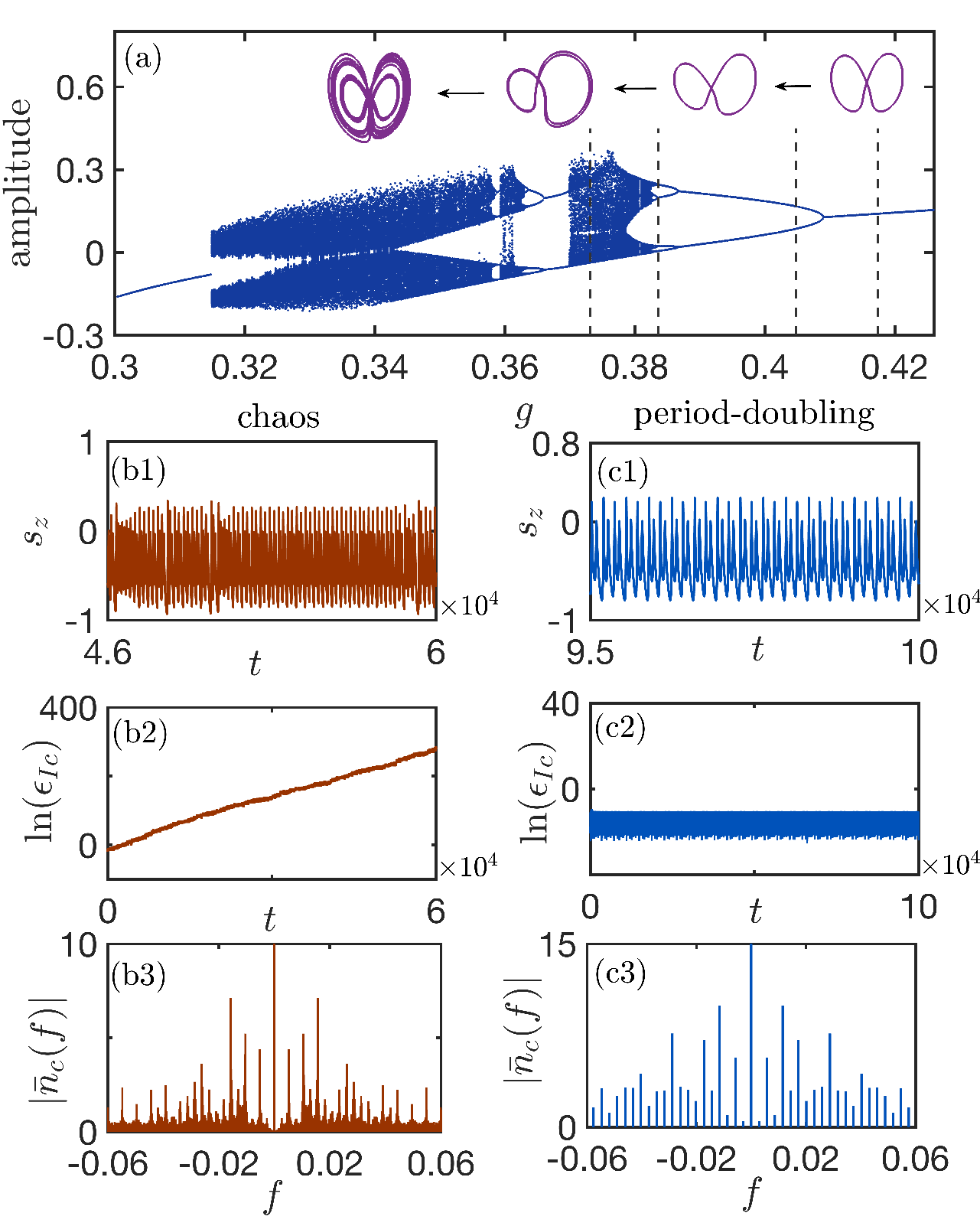}
\caption{\label{fig2} Evolution from periodic to chaotic motion through period-doubling bifurcation. Panel (a): bifurcation phase diagram for $\omega_q=0.1$ which plot the oscillation amplitudes of $s_z(t)=2\langle \hat{J}_z\rangle(t)/N$. The insets (from right to left) present the bifurcation of qubit trajectories from the symmetry limit cycle, unsymmetric limit cycle, via period-doubling to chaos. The dashed black lines indicate the values of parameters. Panels (b1)-(b3) describe chaotic dynamics ($g=0.384$) by qubit evolution $s_z(t)$, perturbations of cavity field $\ln(\epsilon_{I_c})$ and power spectrum density of photon field respectively. Panel (c1)-(c3): evolution of period-doubling ($g=0.373$). Other parameters are the same as Fig.~\ref{fig1}. 
}
\end{figure}
%%%%%%%%%%%%%%%%%%%%%%%%%%%%%%%%%%%%%%%%%%%%%%%%%%%%%%%%%%%%%%%%%%%%%%%%%%%

For a small value of qubit frequency, we found period-doubling bifurcation and cascade leading to chaos. Using the bifurcation diagram in Fig.~\ref{fig2}(a), we illustrated detailedly how a limit cycle transformed into a chaotic attractor. It is obviously seen that oscillations give a single amplitude at large g, while gradually bifurcating to $2^n$ $(n=1,2,3,\dots)$ amplitudes as g decreases. As a typical route to chaos, period-doubling bifurcation has been observed in the anisotropic two-photon Dicke model. Differently, the bifurcation of the periodic orbit here begins with a symmetric-broken limit cycle. For large coupling, qubit trajectory in phase space first exhibits a limit cycle with two symmetric loops, displaying one amplitude in the bifurcation diagram, which consists of the amplitudes of $s_z(t)$. To simplify the observation of bifurcation, we hide the axes of $s_{x,y,z}$ in the insets of qubit trajectories. The two loops become unsymmetric with the decreasing coupling strength, presenting two amplitudes. When the coupling strength decreases further, it splits into a new one with four loops, resulting in quadruple amplitudes in the phase diagram. The corresponding dynamics are depicted in picture (c), where the system evolution shows four amplitudes with fourth periods of the original simple cyclic motion. Dynamics in a regular orbit provides a flat form in the evolution of cavity field perturbation $\ln(\epsilon_{Ic})$, the slope of which determines the value of the Lyapunov exponent (LE). Several discrete peaks can be found in the power spectral density (PSD) of photon number $\bar{n}_c=\langle \hat{a}^{\dagger}\hat{a} \rangle/N$. They are equally spaced in the figure, indicating that the frequency ratios are rational. By further decreasing $g$, cascading of bifurcation occurs. The system finally enters the chaotic regime with infinite amplitudes. Picture (b) depicts chaotic dynamics, which presents a random oscillation with variable amplitude and a continuous spectrum. In phase space, even infinitesimally nearby trajectories will rapidly deviate from one another, leading to an exponential growth of the perturbation and a positive LE.

%%%%%%%%%%%%%%%%%%%%%%%%%%%%%%%%%%%%%%%%%%%%%%%%%%%%%%%%%%%%%%%%%%%%%%%%%%%
\subsection{ Intermittency }
%%%%%%%%%%%%%%%%%%%%%%%%%%%%%%%%%%%%%%%%%%%%%%%%%%%%%%%%%%%%%%%%%%%%%%%%%%%
\label{IC}

%%%%%%%%%%%%%%%%%%%%%============fig3===========================%%%%%%%%%%%%%%%%%%%
\begin{figure*}
\centering
\includegraphics[width=0.95\textwidth]{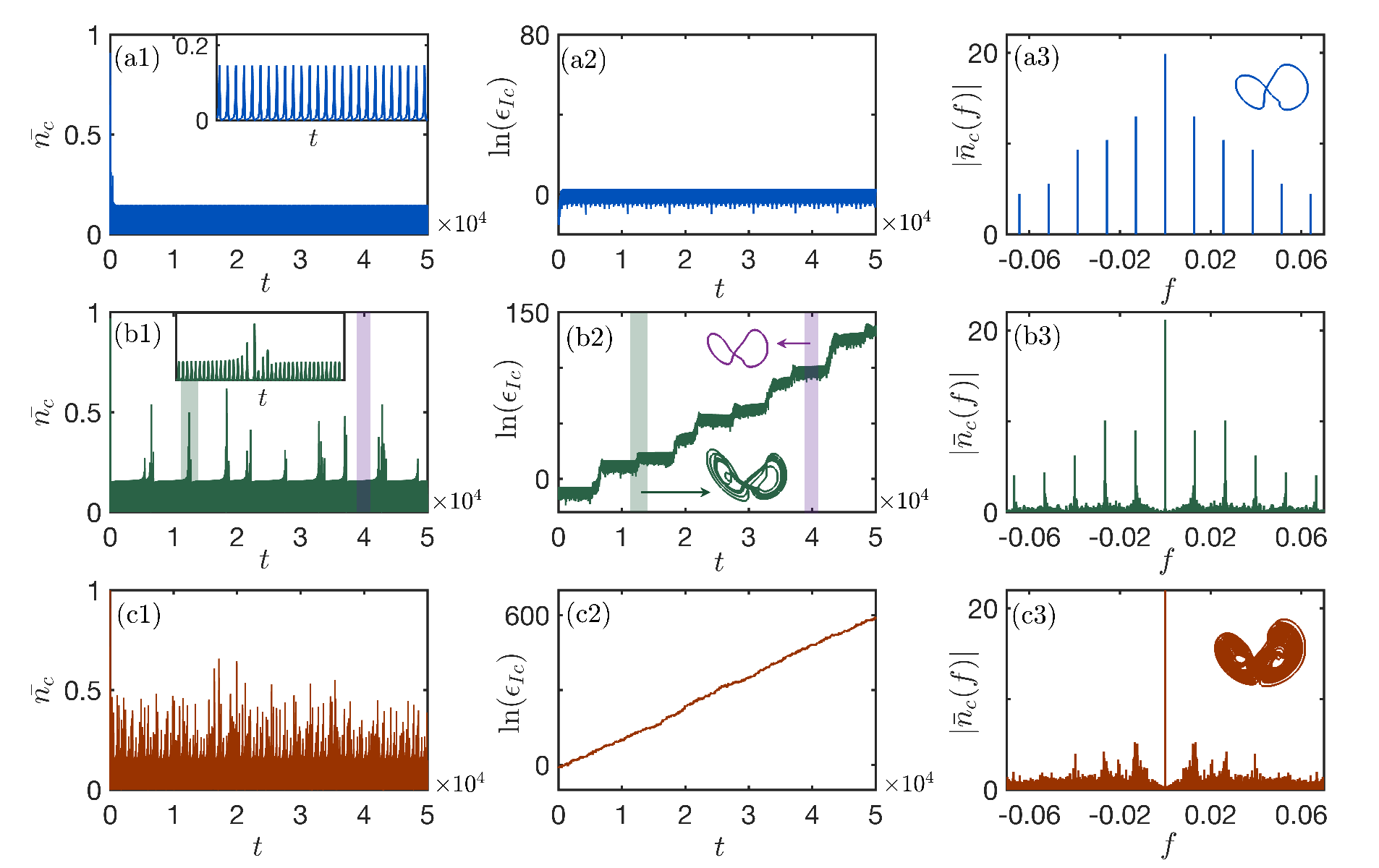}
\caption{\label{fig3} 
Evolution from periodic to fully chaotic motion via intermittent chaos (IC). By plotting the time evolution of photon field $\bar{n}_c=\langle \hat{a}^{\dagger}\hat{a} \rangle/N$, perturbation, and photon field spectrum in the columns from left to right, we present periodic evolution ($g=0.3763$), intermittent chaos ($g=0.3795$), and fully chaotic dynamics ($g=0.3826$) in pictures (a)-(c) respectively. For better insight, we enlarge part of intermittency dynamics ( green shaded ) in the inset of panels (b1). Correspondingly, insets in (b2) display system trajectories of chaotic (green) and periodic sectors (purple) in IC. A medium value of qubit frequency $\omega_q=0.15$ is used in the calculation. Other parameters are the same as Fig.~\ref{fig1}.
}
\end{figure*}

%%%%%%%%%%%%%%%%%%%%%%%%%%%%%%%%%%%%%%%%%%%%%%%%%%%%%%%%%%%%%%%%%%%%%%%%%%%
We discovered that chaos can emerge through intermittent motion when system settings are changed for the medium value of qubit frequency. Intermittency describes dynamics switching between different types of oscillations under fixed control parameters. The switching appears to occur randomly, usually between regular and chaotic behaviors \cite{hilborn2000chaos,zhang2020intermittent}. The detailed process of how chaos appears via intermittent chaos (IC) is exemplified in Fig.~\ref{fig3}, where $\omega_q=0.15$ is adopted.

For small coupling strength, the system evolves on a periodic orbit that is formed by the fixed point $\rm{SP}$ via Hopf bifurcation. Stable oscillation with constant amplitude is shown in Fig.~\ref{fig3}(a). As stated in the previous section, periodic dynamics offer a flat shape in the evolution of perturbation $\ln(\epsilon_{Ic})$ and equidistant discrete peaks in the power spectrum. With increasing coupling strength, occasional bursts of irregular motion in periodic dominant evolution occur. It can be seen from Fig.~\ref{fig3}(b) that, along with time, system behavior switches between period and chaos for constant system parameters, giving rise to a ladder form in the evolution of $\ln(\epsilon_{Ic})$. In panel (b2), short-lived chaos generates the leap part, presenting a strange attractor in phase space (inset plotted by the green line). The flat part corresponds to periodic oscillation, which produces a limit cycle (inset plot by the purple line). During the evolution, system trajectory vacillates between periodic orbits and chaotic attractors. This particular intermittence feature is also demonstrated clearly in the PSD of the cavity field, which distinguishes several equally spaced discrete peaks from numerous small sidebands. Occasional chaotic motion becomes more frequent and lasts longer with further increasing coupling strength. Eventually, the system enters a fully chaotic phase, displaying irregular oscillation, a sloped version of $\ln(\epsilon_{Ic})(t)$, and continuous spectrum, see Fig.~\ref{fig3}(c). In comparison with periodic and intermittent spectra, no dominant frequency can be recognized from the chaotic continuous spectrum. 

%%%%%%%%%%%%%%%%%%%%%%%%%%%%%%%%%%%%%%%%%%%%%%%%%%%%%%%%%%%%%%%%%%%%%%%%%%%
\subsection{ Quasi-Periodicity }
%%%%%%%%%%%%%%%%%%%%%%%%%%%%%%%%%%%%%%%%%%%%%%%%%%%%%%%%%%%%%%%%%%%%%%%%%%%
\label{q-period}

%%%%%%%%%%%%%%%%%%%%============fig4===========================%%%%%%%%%%%%%%%%%%%
\begin{figure}
\centering
\includegraphics[width=0.47\textwidth]{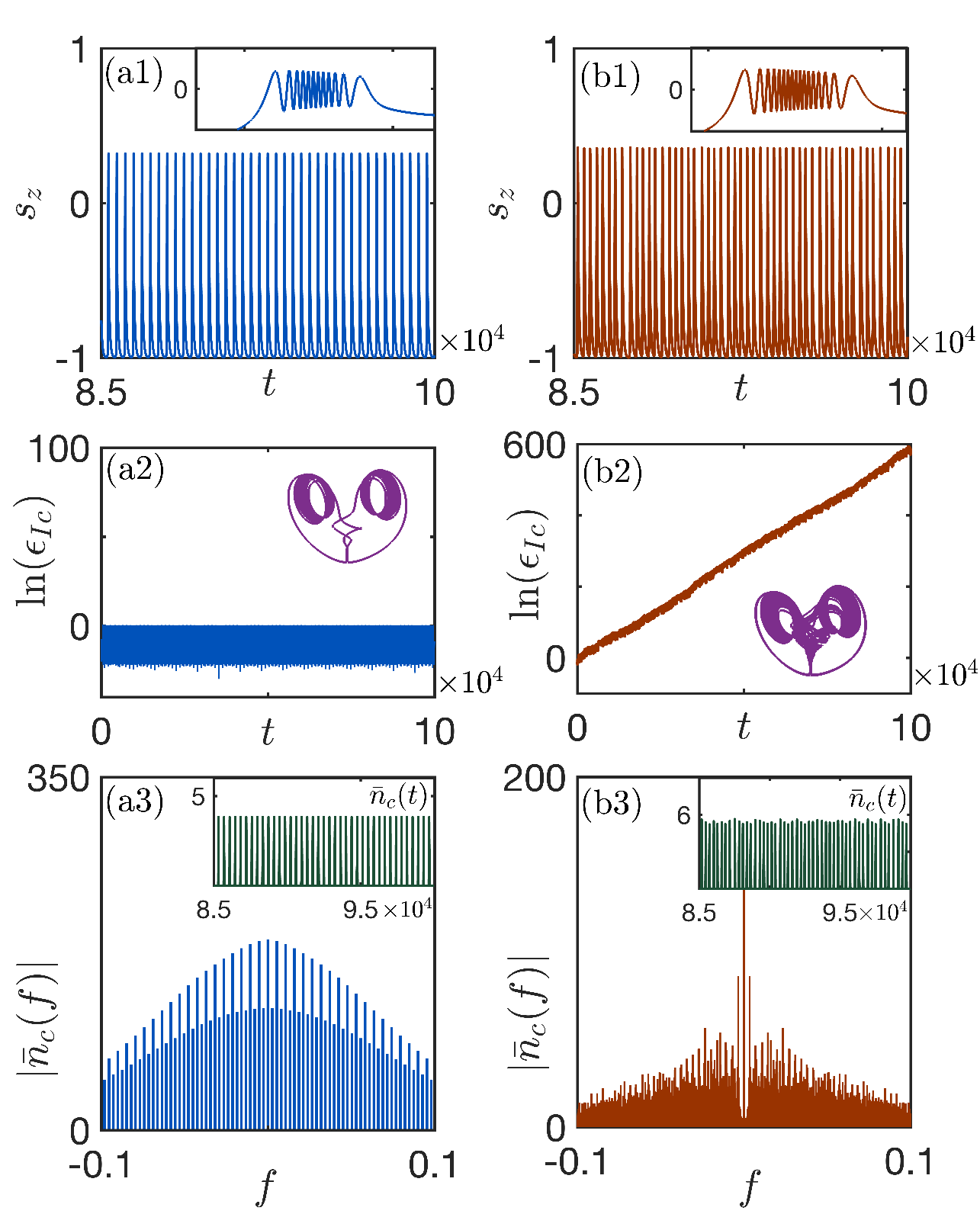}
\caption{\label{fig4} Evolution from quasi-periodic (left panel: $g=0.3716$ ) to chaotic motion (right panel: $g=0.3795$ ) for large qubit frequency $\omega_q=0.25$. The inset in panels (a1) and (b1) is the blowup of a random prominent peak of $s_z(t)$. In the discrete spectrum (panel (a3)), the main frequencies are at integer multiples of $f_1=5.3\times 10^{-3}$ (in units of $\omega_0$), and the distance between auxiliary frequency peaks is $f_2=2.6\times 10^{-4}$. Inset in (b2): chaotic attractor. Insets in (a3) and (b3): evolution of cavity field $\bar{n}_c=\langle \hat{a}^{\dagger}\hat{a} \rangle/N$. Other parameters are the same as in Fig.~\ref{fig1}. 
}
\end{figure}
%%%%%%%%%%%%%%%%%%%%%%%%%%%%%%%%%%%%%%%%%%%%%%%%%%%%%%%%%%%%%%%%%%%%%%%%%%%

%%%%%%%%%%%%%%%%%%%%============fig5===========================%%%%%%%%%%%%%%%%%%%
\begin{figure}
\centering
\includegraphics[width=0.47\textwidth]{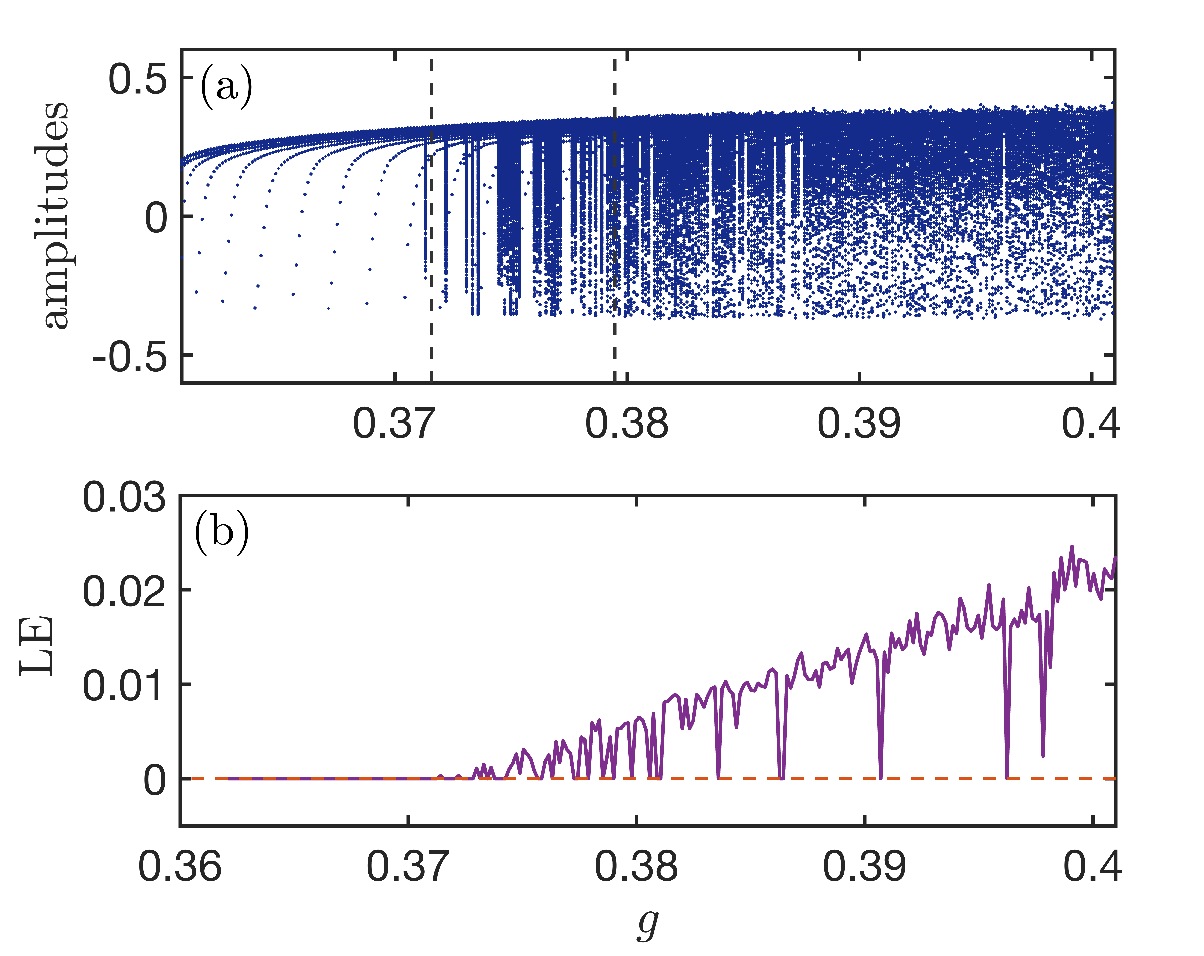}
\caption{\label{fig5} Panel (a): amplitudes statistics of qubit oscillation $s_z(t)$ during quasi-periodicity to chaos. The dashed lines mark the quasi-periodic and chaotic dynamics presented in Fig.~\ref{fig4}. Panel (b): corresponding values of the Lyapunov exponent (LE). The dashed line denotes the zero value. Other parameters are as in Fig.~\ref{fig4}. 
}
\end{figure}
%%%%%%%%%%%%%%%%%%%%%%%%%%%%%%%%%%%%%%%%%%%%%%%%%%%%%%%%%%%%%%%%%%%%%%%%%%%

For large qubit frequency, we found that chaos emerges by way of quasi-periodicity. As can be seen from Fig.~\ref{fig4}(a1), the qubit behaves deceptively with periodic oscillation when the coupling strength is small. Actually, the trajectory never repeats itself exactly. Each prominent peak of the quasi-periodicity includes several additional minor oscillations, presenting a number of amplitudes. Physically, the spins in our system can be regarded as an oscillator that nonlinear couples to another harmonic oscillator -- the bosonic field. The interaction of the two nonlinear oscillators introduces additional frequencies. If the ratio of the frequencies is rational, system dynamics show periodic motion, as we have presented in the previous sections. On the contrary, system behavior shows quasi-periodicity if the ratio is irrational \cite{hilborn2000chaos,patra2019chaotic}. 

In the quasi-periodic scenario, the system trajectory will drift slightly with the change of initial conditions, resulting in a small distance from the original trajectory that will not diverge with time. On the one hand, it produces a flat shape in the evolution of $\ln(\epsilon_{Ic})$ i.e., zero LE. On the other hand, the long time trajectory will cover the surface of a tour in phase space. As can be seen from the inset in Fig.~\ref{fig4}(a2), the trajectory that corresponds to the additional minor oscillations is crammed into two ring surfaces. Under the competition of the two modes, quasi-periodicity creates a distinct power spectrum with periodic oscillation, see Fig.~\ref{fig4}(a3). Though it consists of discrete peaks, the frequencies are not evenly distributed. The dominant peaks are situated at $0, \pm f_1, \pm2 f_1, \dots $, with auxiliary peaks spaced at $f_2$ bunched around them. With increasing coupling strength, each oscillation comprises a growing number of amplitudes. The trajectories from the two tours spread and interweave, finally leading to chaotic motion, seeing in Fig.~\ref{fig4} (b). Correspondingly, positive LE and continuous spectrum also can be seen in the picture.

Similar to the bifurcation diagram in Fig.~\ref{fig2}, we record the amplitudes of $s_z(t)$ during the appearance of chaos through quasi-periodicity. As shown in Fig.~\ref{fig5}(a), both quasi-periodic and chaotic oscillation shows a larger number of peaks, while the former can be counted. When the dynamic of the system gets chaotic, an overwhelming number of amplitudes arise in a narrow band of $g$, producing a dense concentration of points in the picture. The variation of the Lyapunov exponent with coupling strength is shown in Fig.~\ref{fig5}(b), which is in good agreement with the amplitude statistics graph. LE is close to zero in the domain of finite amplitudes, corresponding to quasi-periodic oscillations. Positive LE, as reflected by infinite amplitudes, provides evidence for the existence of chaos.

%%%%%%%%%%%%%%%%%%%%%%%%%%%%%%%%%%%%%%%%%%%%%%%%%%%%%%%%%%%%%%%%%%%%%%%%%%%
\section{overall phase diagram}
%%%%%%%%%%%%%%%%%%%%%%%%%%%%%%%%%%%%%%%%%%%%%%%%%%%%%%%%%%%%%%%%%%%%%%%%%%%
\label{phase}

%%%%%%%%%%%%%%%%%%%%============fig6===========================%%%%%%%%%%%%%%%%%%%
\begin{figure}[t]
\centering
\includegraphics[width=0.45\textwidth]{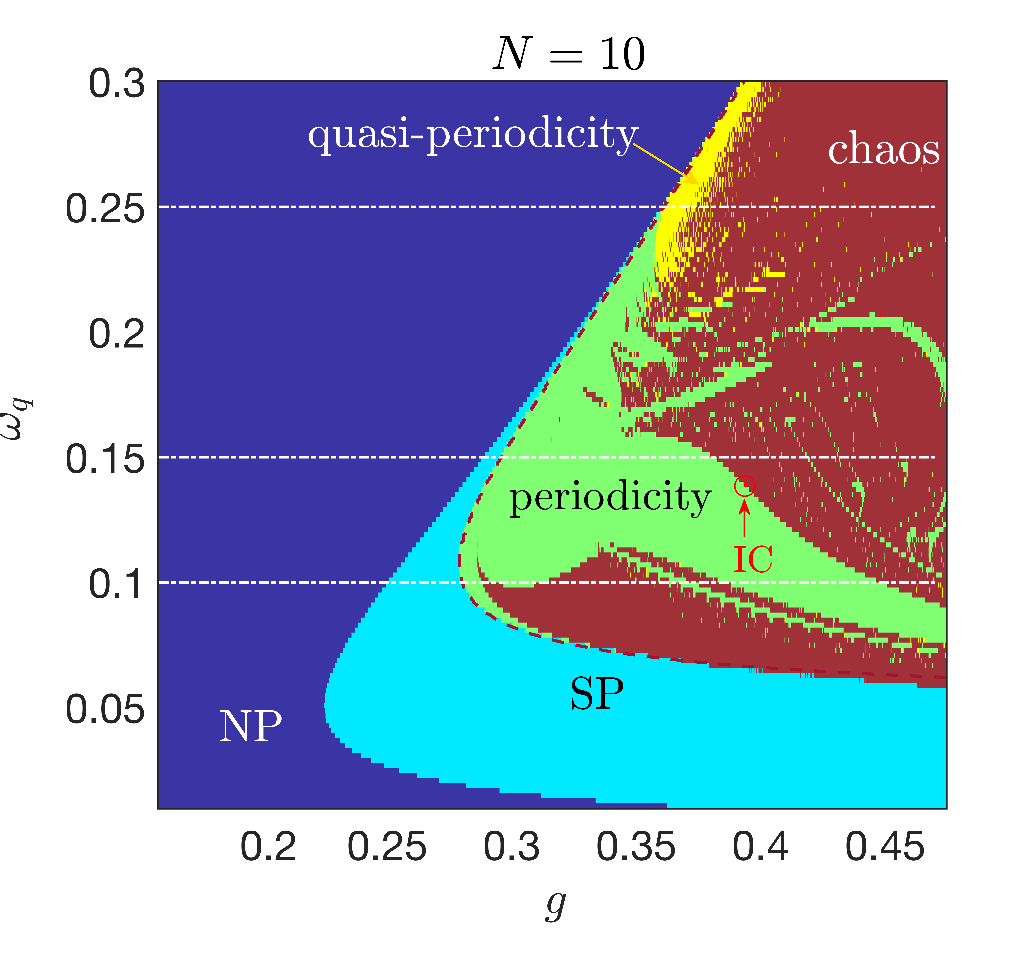}
\caption{\label{fig6} Overall phase diagram. It contains stable phases -- $\rm{NP}$ (dark blue), $\rm{SP}$ (light blue), and oscillatory phases -- periodic (green), quasi-periodic (yellow), intermittent chaos (red spots), full chaos (broown). The dashed-dotted lines marked the values of qubit frequencies that we have used in previous figures. The red dashed line is the critical boundary $g_{t2}$. Other parameters are the same as in Fig.~\ref{fig1}. 
}
\end{figure}
%%%%%%%%%%%%%%%%%%%%%%%%%%%%%%%%%%%%%%%%%%%%%%%%%%%%%%%%%%%%%%%%%%%%%%%%%%%

The emergence of chaos in various ways results from the nonlinear interaction between the spin and cavity field. The inclusion of qubit dissipations enables us to treat spin as an independent oscillator whose dynamics will not be constrained on the Bloch sphere's surface. Including stable fixed points and chaotic dynamics, we provide an overall phase diagram that changes with coupling strength $g$ and qubit frequency in Fig.~\ref{fig6}. The stable phases $\rm{NP}$ and $\rm{SP}$ are indicated by dark and light blue areas in the figure. The unstable phase in Fig.~\ref{fig1} now is replaced by periodicity (green), quasi-periodicity (yellow), intermittent chaos (red), and full chaotic (brown) phases. When calculating the phase diagram, it is useful to distinguish periodic motion and quasi-periodicity from intermittency and chaos using the Lyapunov exponent, where regular oscillations yield zero values and chaotic dynamics produce positive values. Consequently, we need to differentiate quasi-periodicity from the period. Both of them have discrete spectra and zero LE. We can recognize quasi-periodicity by its non-repetitive and irregular time evolution. When compared to periodic oscillation, significantly increased amplitudes could be discovered in quasi-periodicity during the same time interval. Moreover, the amplitudes generated by quasi-periodicity will change with oscillation time and sampling frequency. Also, we can identify quasi-periodicity by the power spectrum, which shows an irrational ratio of frequencies. Then we can separate intermittency and full chaos through the ladder form of $\ln(\epsilon_{Ic})(t)$.

As we can see from the figure, the system evolves on a periodic orbit produced by fixed point $\rm{SP}$ via Hopf bifurcation when the coupling strength is slightly large than $g_{t2}$. Due to the competition between the spin and bosonic modes, an additional frequency may arise as the coupling strength increases. The dashed-dotted lines indicate the position of the three typical qubit frequencies which have been used in Figs.~\ref{fig2}-\ref{fig5}. For small qubit frequency, which is represented by $\omega_q=0.1$ (bottom line), one mode is locked by another one under strong coupling, leading to period-doubling dynamics over a finite control parameter regime. The output light will display intense bursts or pulses, which is a very popular way of generating frequency comb. By properly adjusting system parameters, the two modes exhibit balanced influences on the system, which gives rise to intermittent chaos. The parameter we used in Fig.~\ref{fig3}, which show the process of intermittent chaos to full chaos, are chosen along the middle line $\omega_q=0.15$. IC is uncommon in our system, it only shows several spots in the picture. For large qubit frequency, the frequency induced by the two competition modes will show a slight deviation from the original frequency. Incommensurate frequencies lead to irregular oscillation and quasi-periodicity. Before approaching the quasi-periodic state, the system usually experiences a periodic phase that diminishes with increasing qubit frequency.

Now we can compare the dynamics of the current model to the anisotropic two-photon Dicke model that only considers cavity field dissipation \cite{li2022nonlinear}. Aside from the differences highlighted in the phase diagram of stable fixed points in Sec.~\ref{model}, the presented model also exhibits distinct nonlinear dynamics. In Ref.~\cite{li2022nonlinear}, chaos only can be found in an anisotropic condition ($\lambda>\lambda_t>1$), which appears through the cascade of period-doubling bifurcations. For a given set of parameters, multiple types of motion coexist in phase space, resulting in an impressionable phase diagram with system initial states. The current work, which is implemented in an isotropic instance $\lambda=1$, likewise shows the sign of chaos. Expect the cascade of period-doubling bifurcations, chaos also can occur via intermittency and quasi-periodicity by adjusting qubit frequency, creating a colorful phase diagram. Since there is no coexistence of different motions, the phase diagram will not be influenced by the system's initial state. Other chaos-related dynamics, such as period window and collision of chaotic attractors, can also be discovered in the current model and are not discussed here in detail.

%%%%%%%%%%%%%%%%%%%%%%%%%%%%%%%%%%%%%%%%%%%%%%%%%%%%%%%%%%%%%%%%%%%%%%%%%%%
\section{conclusion}
%%%%%%%%%%%%%%%%%%%%%%%%%%%%%%%%%%%%%%%%%%%%%%%%%%%%%%%%%%%%%%%%%%%%%%%%%%%
\label{conclusion}

In conclusion, we investigated the semi-classical nonlinear dynamics of the two-photon Dicke model by considering different dissipation channels. The qubit decay and dephasing help to stabilize the system, where the localized phase that reflects the spectral collapse in Hamiltonian format and coexistence phases vanishes. Under balanced rotating and counter-rotating couplings, alongside the normal and superradiant phases, a phase encompassing a variety of chaos-related phenomena, such as period doubling, quasi-periodic oscillation, and intermittent chaos, is established. More interestingly, by adjusting qubit frequency, we found that chaos can emerge in three different ways. We provide a comprehensive illustration of the transitions using bifurcation diagrams and a global chaotic phase diagram. Comparisons of our findings to prior research in the field also have been made.

It might be possible to realize the dissipative two-photon Dicke model in driven systems such as trapped ion chains and quantum superconducting circuits, which can eliminate undesirable effects caused by ultra-strong coupling in loss-dominated systems~\cite{felicetti2015spectral,lv2018quantum,puebla2017protected,felicetti2018ultrastrong,de2018breakdown}. Under effective implementations of the model, coupling strength can be comparable to (or even larger than) the bosonic and atomic (effective) frequencies, as well as considered decoherence and dissipation processes. The dissipation channels can be tailored in strongly driven atomic cloud or quantum simulation systems using bath-engineering approaches~\cite{poyatos1996quantum,verstraete2009quantum,baumann2010dicke,liu2011light,asjad2014reservoir,kirton2017suppressing}.

The present discussion considers the semiclassical limit of the collective spin variables, but it would be also of great interest to investigate how these nonlinear phenomena manifest themselves in fully quantum treatments~\cite{wimberger2014nonlinear,PhysRevLett.61.1899,PhysRevLett.123.254101}. Studies have shown that signatures of dissipative quantum chaos can be detected by the Liouvillian superoperator, whose spectrum shows unique statistical properties~\cite{PhysRevX.10.021019,PhysRevB.101.214302,PhysRevB.101.214302,PhysRevA.102.033715,thingna2021degenerated,PhysRevResearch.3.043197,li2021spectral,yusipov2022quantum,lange2021random}. Furthermore, the anisotropic coupling may also be included in the present dissipative Dicke model, which would likely result in more complicated dynamics and phase diagrams. Our investigation into the  nonlinear dynamics of this dissipative quantum systems may stimulate further research on quantum chaos, the findings of which may be applicable to quantum on-chip devices and communications.

%%%%%%%%%%%%%%%%%%%%%%%%%%%%%%%%%%%%%%%%%%%%%%%%%%%%%%%%%%%%%%%%%%%%%%%%%%%
\section{Acknowledgments}
%%%%%%%%%%%%%%%%%%%%%%%%%%%%%%%%%%%%%%%%%%%%%%%%%%%%%%%%%%%%%%%%%%%%%%%%%%%

We gratefully acknowledge R. Fazio for his helpful discussions. S.C. acknowledges support from the Innovation Program for Quantum Science and Technology (Grant No. 2021ZD0301602), the National Science Association Funds (Grant No. U2230402), and the National Natural Science Foundation of China (Grant Nos. 11974040 and 12150610464).

%%%%%%%%%%%%%%%%%%%%%%%%%%%%%%%%%%%%%%%%%%%%%%%%%%%%%%%%%%%%%%%%%%%%%%%%%%%

\bibliographystyle{apsrev4-2}
\bibliography{tp_qubitdecay}

\end{document}